\documentclass[11pt]{article}

\usepackage{amsmath}
\usepackage[dvips]{graphicx}

\usepackage{natbib}

\title{Microwave emissivity of fresh water ice\\--Lake ice and Antarctic ice pack--\\Radiative transfer simulations versus satellite radiances.}

\author{Peter Mills\\Peteysoft Foundation\\\textit{petey@peteysoft.org}}

\begin{document}

\maketitle

\begin{abstract}
Microwave emissivity models of sea ice are poorly validated empirically.
Typical validation studies involve using averaged or stereotyped profiles of
ice parameters against averaged radiance measurements.
Measurement sites are rarely matched and even less often point-by-point.
Because of saline content, complex permittivity of sea ice is highly variable
and difficult to predict.
Therefore, to check the validity of 
a typical, plane-parallel, radiative-transfer-based ice emissivity model, 
we apply it to fresh water ice instead of salt-water ice.
Radiance simulations for lake ice are compared with
measurements over Lake Superior 
from the Advanced Microwave Scanning Radiometer on EOS (AMSR-E).
AMSR-E measurements are also collected over Antarctic icepack.
For each pixel, a thermodynamic model is driven by four years of 
European Center for Medium Range Weather Forecasts (ECMWF) reanalysis data
and the resulting temperature profiles used to drive the emissivity
model.
The results suggest that the relatively simple emissivity model is a good
fit to the data.
Both cases, however, show large discrepencies whose most likely explanation
is scattering both within the ice sheet as well as by cloudy atmospheres.
Scattering is neglected by the model.
Further work is needed to refine the scattering component of ice emissivity
models 
and to generate accurate estimates of complex permittivities within sea ice.
\end{abstract}

\section{Introduction}

In \citet{Mills_Heygster2009}, 
a plane-parallel radiative transfer (RT) model was
used to simulate emitted microwave brightness temperatures at L-band
(1.4 GHz) frequencies.  The question addressed in this paper is 
simply the following: is such a model a reasonable approximation to
reality?  Can factors not accounted for in the model, such as ice ridging
and surface scattering, be neglected?  The results of \citet{Mills_Heygster2009}
suggest that while the effect of ice ridging is small, it is not insignificant.

Because of the simplicity of the model and because effective permittivities,
required as input to the model, are notoriously difficult to estimate
for saline ice \citep{Ulaby_etal1986,Mills_Heygster2009}, 
such a model would be a good candidate for inverse methods
that aim to retrieve physical properties of sea ice beyond mere concentration.
The idea is to retrieve those quantities most relevant to the model,
namely the complex permittivities, rather than going one step further back
and trying to retrieve ice bulk and microstructural properties.
The model is only appropriate to lower (below 25 GHz) frequencies where 
volume scattering is weak 
\citep{Mills_Heygster2009,Barber_etal1998,Vant_etal1978,Stogryn1986,Johnsen1998}.  
The functional dependence of permittivity with frequency 
in this range also tends to be weak \citep{Vant_etal1978}, 
thus it may be possible to assume constant permittivity across all frequencies.

With the launch of the new SMOS instrument operating at L-band, the
possibilities for such a retrieval expand considerably.  With SMOS, there
is some potential to retrieve ice thickness.  In the event that ice is too
opaque for this, however, another good candidate for retrieval is 
snow-thickness.  Dry snow is nearly transparent at L-band but becomes 
progressively more opaque at higher frequencies (see Equation (\ref{att_coef})).

The motivation for this study is two-fold:  first, there is a near absence
of direct, point-by-point validation studies of ice emissivity models of
this type, that is, by taking ice cores, feeding the measured profiles
into an emissivity model and comparing
the results to radiance measurements taken at corresponding locations.
Most validation studies consist of comparing averaged measured brightness
temperatures with simulations based on averaged or idealized vertical
ice profiles.  Second, the complex permittivity of pure ice is much
less variable than that of saline ice, making it easier to predict.
Because of the very low value of the imaginary part, which determines
the attenuation, 
the effect of ice thickness on the signal over thin ice, i.e., over
lake ice, will be relatively small.

The procedure will be to simulate microwave emission over freshwater ice--
Lake Superior and the Antarctic ice cap--and compare these with satellite
measurements from the Advanced Microwave Scanning Radiometer on EOS (AMSR-E).

\section{Models}

\subsection{Emissivity model}

The radiative transfer model used here is described thoroughly
in \citet{Mills_Heygster2009} so only the briefest treatment will be given
here.  A plane parallel geometry is assumed making it possible to solve for upwelling
and downwelling radiances along a single line-of-sight:
\begin{align}
T_i \uparrow - \tau_i (1-R_i) T_{i+1} \uparrow - \tau_i R_i T_i \downarrow 
& = (1 - \tau_i) T_i 
\label{rt1}\\
T_i \downarrow - \tau_i (1-R_{i-1}) T_{i-1} \downarrow - \tau_i R_{i-1} T_i \uparrow 
& = (1 - \tau_i) T_i,
\label{rt2}
\end{align}
where $T_i$ is the physical temperature of the $i$th layer, 
$T_i\uparrow$ and $T_i\downarrow$ are the upwelling and downwelling radiances
(as brightness temperatures) respectively.  If there are discontinuous interfaces
between the layers, the reflection coefficients, $R_i$ can be calculated from
the Fresnel equations.  The transmission coefficients, $\tau_i$, is derived:
\begin{equation}
\tau_i = \exp \left (- \frac{\alpha_i \Delta z_i}{\cos \theta_i} \right ),
\end{equation}
from the layer thickness, $\Delta z$, the transmission angle, $\theta_i$
(calculated from Snell's law), and the attenuation coefficient, $\alpha_i$:
\begin{equation}
\alpha_i = \frac{4 \pi \nu}{c} \mathrm{imag} n_i,
\label{att_coef}
\end{equation}
which, in turn, depends on the frequency, $\nu$ and the imaginary part of the refractive
index, $n_i=\sqrt{\epsilon_i}$.  $c$ is the speed of light and $\epsilon_i$ is the
relative permittivity.

To run the model, we will need relative permittivities as input.  The real
part for pure ice tends to be fairly constant at around 3.15.  
The following, temperature-dependent model is given in \citet{Maetzler2006}:
\begin{equation}
\epsilon_{pi}^\prime=3.1884+9.1\times10^{-4} T
\end{equation}
where $T$ is the physical temperature in degrees Celsius.
Ice is almost a perfect dielectric: 
the imaginary permittivity at microwave frequencies is very small, 
generally less than 0.01.  We use the model from \citet{Hufford1991}
to model the imaginary permittivity of pure ice as a function of
temperature and frequency.  These estimates will suffice for lake ice,
however the density of ice pack in the Antarctic varies between less
than 500 kg/$\mathrm m^3$ at the surface to over 900 kg/$\mathrm m^3$ or close to the
density of pure ice, at tens of metres depth.  Since the ice pack is
typically granular, we use a mixture model for spherical inclusions
to calculate the effective permittivity \citep{Sihvola_Kong1988}:
\begin{equation}
\epsilon^* = \epsilon_1 + \frac{3 f (\epsilon_2-\epsilon_1) \epsilon_1/(\epsilon_1+2\epsilon_2)}
		{1-f (\epsilon_2-\epsilon_1)/(\epsilon_2+\epsilon_1)}
\end{equation}
where $\epsilon_1$ is the complex permittivity of pure ice,
$\epsilon_2=1$ is the permittivity of air and $f=1-\rho/\rho_{pi}$
is the relative volume of air, which we can calculate from the density
of the snowpack, $\rho$, versus the density of pure ice, $\rho_{pi}$.

We use the following exponential model from \citet{Rist_etal2002}
for the density of snowpack as a function of depth:
\begin{equation}
\rho=918.-539. \exp(-z/32.5)
\end{equation}
where $z$ is depth.

\subsection{Thermodynamic model}

\label{thermo_model}

For the 6GHz channel, 
the most important determinant of the microwave signature of Antarctic 
ice pack is the temperature
and our results will show this.
For higher frequencies, the temperature becomes less important, but
is still nonetheless significant.
Because of the high penetration depth of low-frequency microwaves
in pure ice, temperatures quite deep in the ice will affect the
signal.  They will also vary quite significantly from those on the
surface because of thermal conduction lag.  Thermal conduction is
modelled in one dimension with the diffusion equation:
\begin{equation}
\frac{\partial T}{\partial t} = \frac{\rho_0 \kappa}{\rho C_p} \frac{\partial^2 T}{\partial z^2}
\end{equation}
where $\kappa$ is the thermal conductivity of ice, $C_p$ is its
heat capacity and $\rho_0$ its maximum density.  
The top layer is forced from European Centre
for Medium Range Forecasts (ECMWF) fields of surface air temperature, $T_a$,
surface wind speed, $v$, humidity, $q$, and cloud-cover, $c$.  
These are used to determine the surface heat
flux, $Q^*$, which is a sum of the four components of the flux:
\begin{equation}
Q^*=
Q_{SW} (t, \phi, c) + Q_{LW} (T_a^4, T_0^4, c) + 
	Q_E (e_s(T_a), e_s(T_0), v) + Q_H (T_a, T_0, v)
\end{equation}
which are, from left to right: shortwave, longwave, latent and sensible.
The functional dependencies for typical parameterizations are given
(see, for instance \citet{Yu_Lindsay2003}); $T_0$ is the surface temperature
and $e_s$ is the saturation vapour pressure.  
The shortwave flux is calculated primarily from geometrical considerations.
The local solar zenith angle is calculated as follows:
\begin{equation}
\beta = \cos^-1 \left ( \frac{\cos \phi \cos \theta \cos \lambda - 
		\sin \phi \cos \delta \sin \lambda}
		{\sqrt{\cos^2 \delta \sin^2 \lambda + \cos^2 \lambda}} \right )
\end{equation}
where $\phi$ is the latitude, $\lambda$ is the Earth's tilt,
$\theta$ is the time of day as an angle (angle of rotation of the Earth
around its axis) with $\theta=0$ 
representing noon while $\theta=\pi$ is midnight, 
$\delta$ is the day of the year (angle of revolution 
about the sun) with $\delta=0$ being the winter solstice.

To return the flux, we multiply the cosine of the local solar zenith angle
with the solar ``constant,'' $S$, the cloud cover and the albedo, $a$:
\begin{equation}
Q_{SW} = S \cos \beta (1 - 0.62 c)(1 - a)
\end{equation}

Spatial derivative were calculated using finite-differencing while the
equation was integrated in time with a fourth-order Runge Kutta \citep{nr_inc2}.
The simulation was initialized with a linear temperature profile, 
starting at freezing at the bottom 300 m which was used as the fixed boundary
condition \citep{Rist_etal2002}.  It was then spun up by repeated forcing with
four years of National Center for Environmental Prediction (NCEP) 
reanalysis data.

\subsection{Atmospheric model}

To correct for atmospheric influences we use the parametrised model from
\citet{Wentz_Meissner2000,Wentz1997}.  In the Wentz-Meissner model, atmospheric
transmissivity and atmospheric components of upwelling and downwelling radiation
are derived from column water-vapour and cloud water path based on fitted
polynomials.  The upwelling and downwelling brightness temperatures 
are modelled as equivalent temperatures assuming an isothermal atmosphere.

The total downwelling radiation from the atmosphere will be given as follows:
\begin{equation}
T_{ba} \downarrow = (1 - \tau_a) T_d + \tau T_{bsky}
\label{atm_corr_down}
\end{equation}
where $\tau_a$ is the atmospheric transmissivity, $T_d$ is the equivalent
atmospheric temperature for the downwelling case and $T_{bsky}$ is the
cosmic background radiation.  The upwelling radiation will be given
as:
\begin{equation}
T_{ba} \uparrow = \tau_a T_1\uparrow + (1-\tau) T_u
\label{atm_corr_up}
\end{equation}
where $T_1\uparrow$ is the modelled surface brightness temperature
from the solution of Equations (\ref{rt1}) and (\ref{rt2}) and $T_u$
is the equivalent atmospheric temperature for the upwelling case.
If the upwelling and downwelling temperatures are equal, $T_d=T_u$,
(they are typically very close) then we can model the atmosphere by
simply adding an extra layer (the atmosphere) to the ice RT model 
(\ref{rt1}) and (\ref{rt2}) and setting the topmost reflection coefficient
(interface between the atmosphere and space) to zero.

\section{Results}

\subsection{Simulation of lake ice}

\begin{figure}
\includegraphics[angle=90,width=0.9\textwidth]{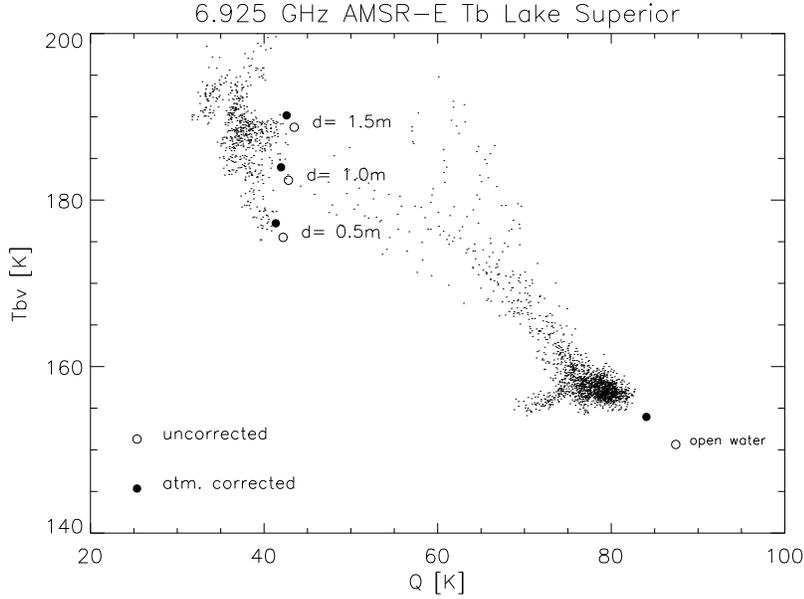}
\caption{AMSR-E brightness temperatures at 6 GHz over Lake Superior during the winter
of 2003.  Dots are measured, circles are model results, both uncorrected (open)
and corrected for atmospheric influences (open circles).}
\label{lake6GHz}
\end{figure}

\begin{table}
\caption{Table of atmospheric correction coefficients calculated according to
\citet{Wentz_Meissner2000}.  
See Equations (\ref{atm_corr_down}) and (\ref{atm_corr_up}) for their use.
Column water vapour was set at 20 kg/$\mathrm{m^2}$
and cloud water path at 0.1 mm.}
\label{atm_coeff}
\begin{center}
\begin{tabular}{|c|c|c|c|}
\hline
$\nu$ [GHz] & $\tau_a$ & $T_u$ [K] & $T_d$ [K] \\
\hline \hline
6.925 & 0.981 & 260.7 & 260.9 \\
10.65 & 0.974 & 262.5 & 262.6 \\
18.70 & 0.911 & 268.8 & 269.3 \\
\hline
\end{tabular}
\end{center}
\end{table}

\begin{figure}
\includegraphics[angle=90,width=0.9\textwidth]{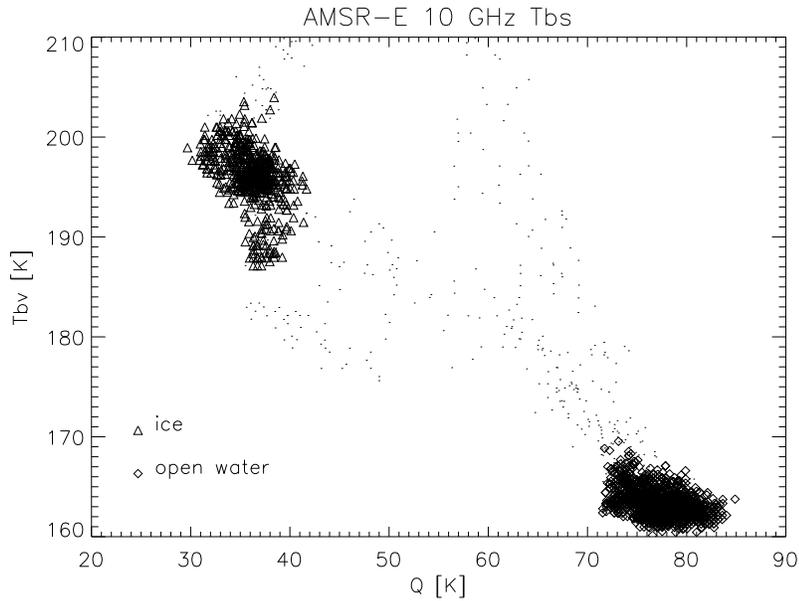}
\caption{Measured AMSR-E brightness temperatures at 10 GHz over Lake Superior during the winter
of 2003.  A clustering algorithm has been used to separate lake ice from open water
points, with the triangles being ice and the diamonds open water.}
\label{cluster_demo}
\end{figure}

\begin{figure}
\includegraphics[angle=90,width=0.9\textwidth]{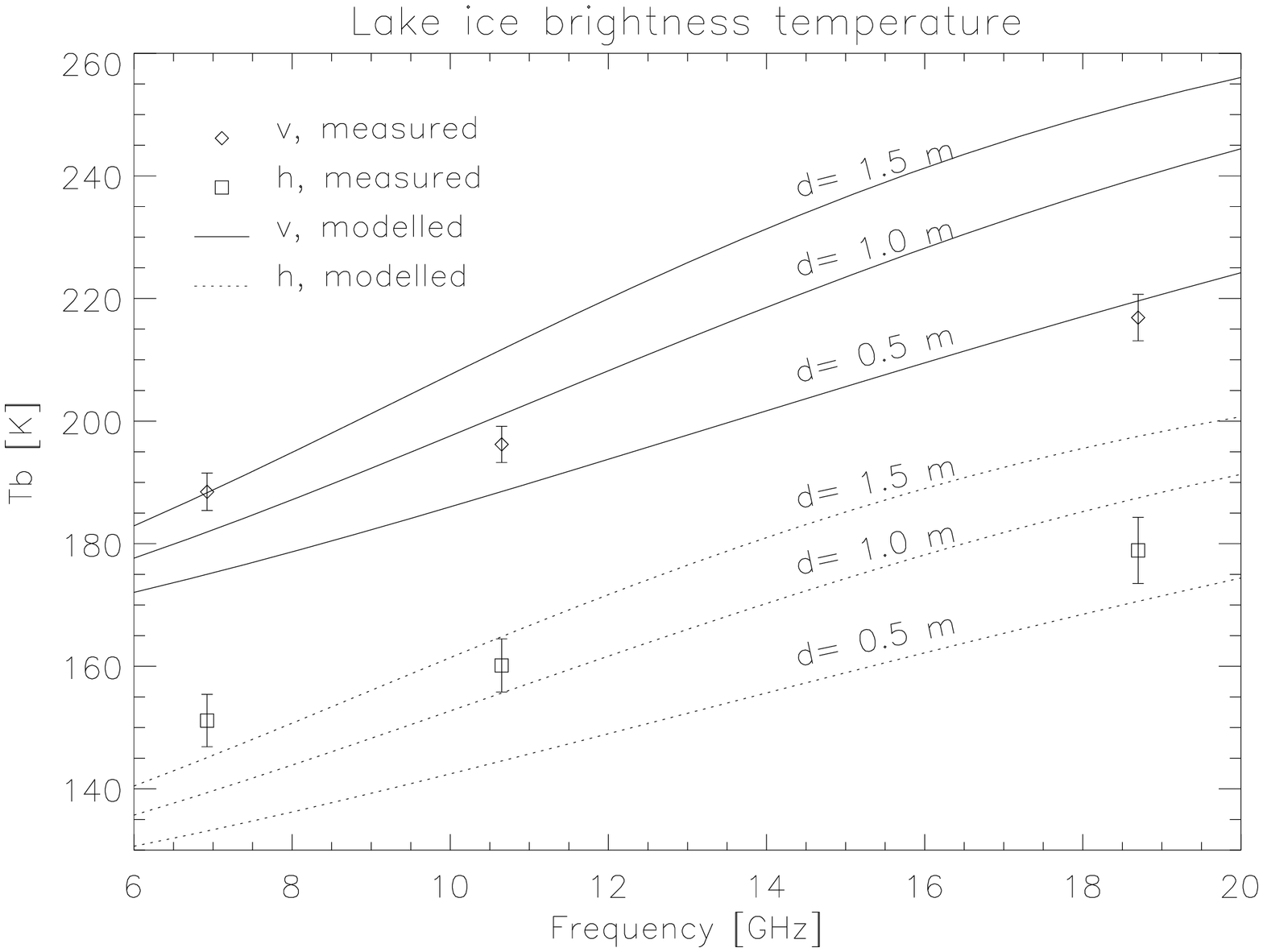}
\includegraphics[angle=90,width=0.9\textwidth]{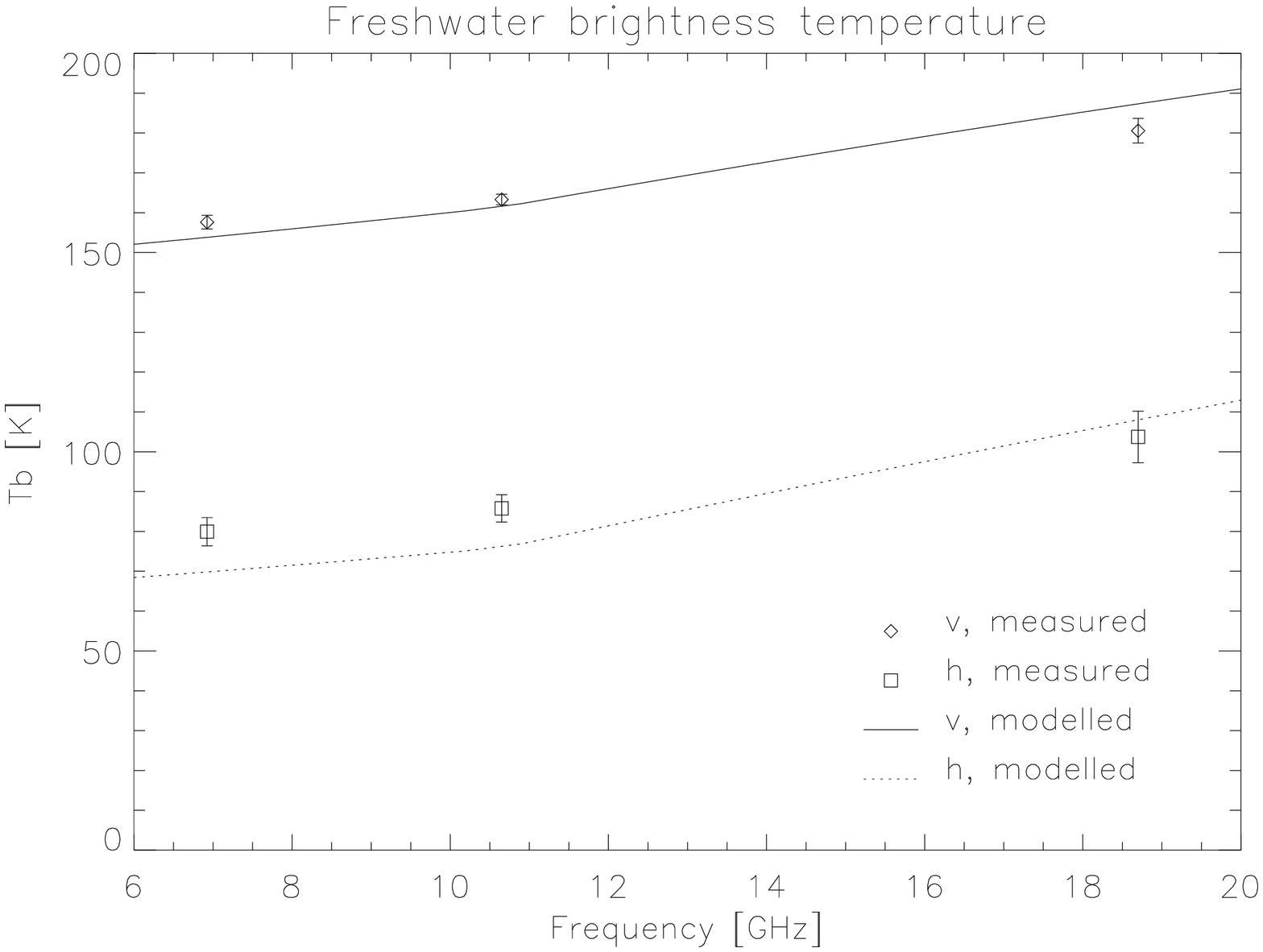}
\caption{AMSR-E brightness temperatures over Lake Superior during the winter
of 2003.  Lines show model results while points with error bars are measured.
Top is ice, bottom is open water.
}
\label{lake_spec}
\end{figure}

AMSR-E measurements were collected over the middle of Lake Superior
(in a radius of 20 km from the coordinate 87 degrees E longitude 
and 48 degrees N latitude) for the winter of 2003 during which there
was much lake ice.  These are compared with brightness temperature
simulations for the 6 GHz case in Figure \ref{lake6GHz}.
A constant ice temperature of zero degrees Celsius was assumed.
The water temperature was likewise assumed constant at freezing
and its complex permittiviy modelled with a Debye relaxation curve
\citep{Ulaby_etal1986}.

Measured and modelled values are compared in a brightness temperature
space consisting of the vertically polarised brightness temperature
versus the polarisation difference.  The cluster of dots on the bottom right
is open water while the cluster on the top left is lake ice
while the larger circles are the model results, both corrected for
atmospheric effects and uncorrected.  Correction coefficients are
shown in Table \ref{atm_coeff}, calculated according to \citet{Wentz_Meissner2000}
assuming total column water vapour of 20 kg/$\mathrm{m^2}$ and
an ice water path of 0.1 mm.

Simulations are performed for three different ice thicknesses.
Fresh water ice is almost a perfect dielectric -- 
because of the low value of the imaginary part of the permittivity,
ice thickness will have less of an effect on the final,
emitted brightness temperature.  Since it is lake ice, we expect
it to be relatively thin, certainly less than 2 m.

We use a clustering algorithm based on kernel density estimation
\citep{Michie_etal1994,Terrell_Scott1992,Mills2009} to group the
open water and ice points and separate the two classes from one another
as demonstrated in Figure \ref{cluster_demo}.  Averages for each cluster,
along with tolerances calculated from the standard deviations are
plotted as a function of frequency in Figure \ref{lake_spec} along
with model results.

\subsection{Simulation of Antarctic icepack}

\begin{figure}
\includegraphics[width=0.8\textwidth]{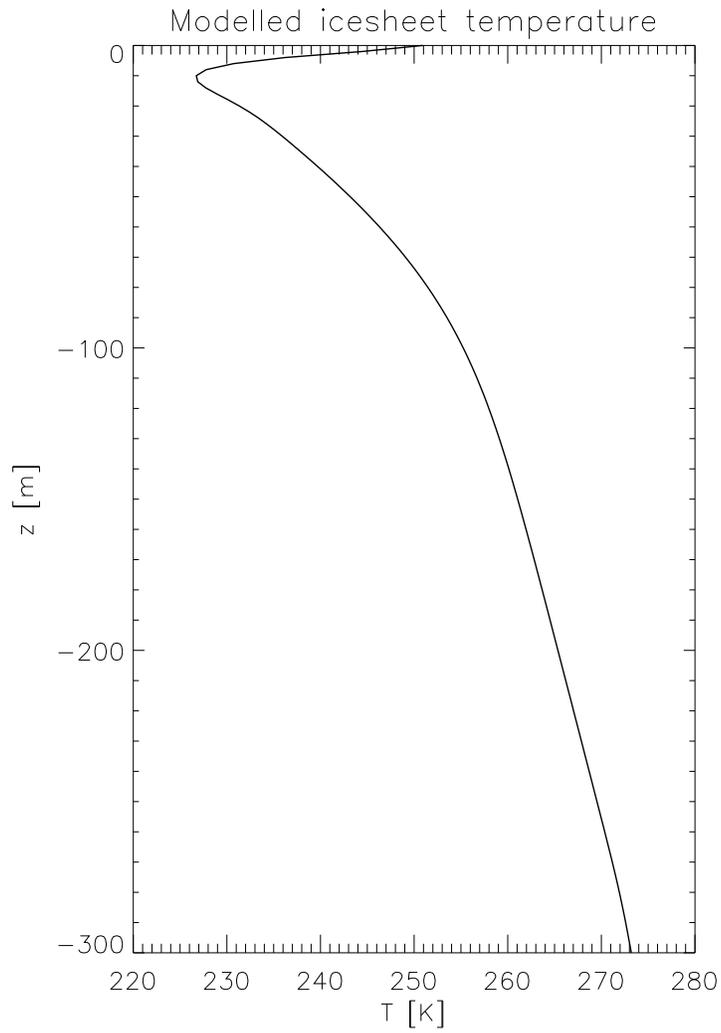}
\caption{Antarctic ice sheet temperature profile for 27 November
2007, 14:42:41 UT at $102.97^\circ$ E, $76.60^\circ$ S.  
Profile has been modelled from four years of
NCEP reanalysis data using a thermodynamic model.}
\label{sample_tprof}
\end{figure}

\begin{figure}
\includegraphics[width=0.9\textwidth]{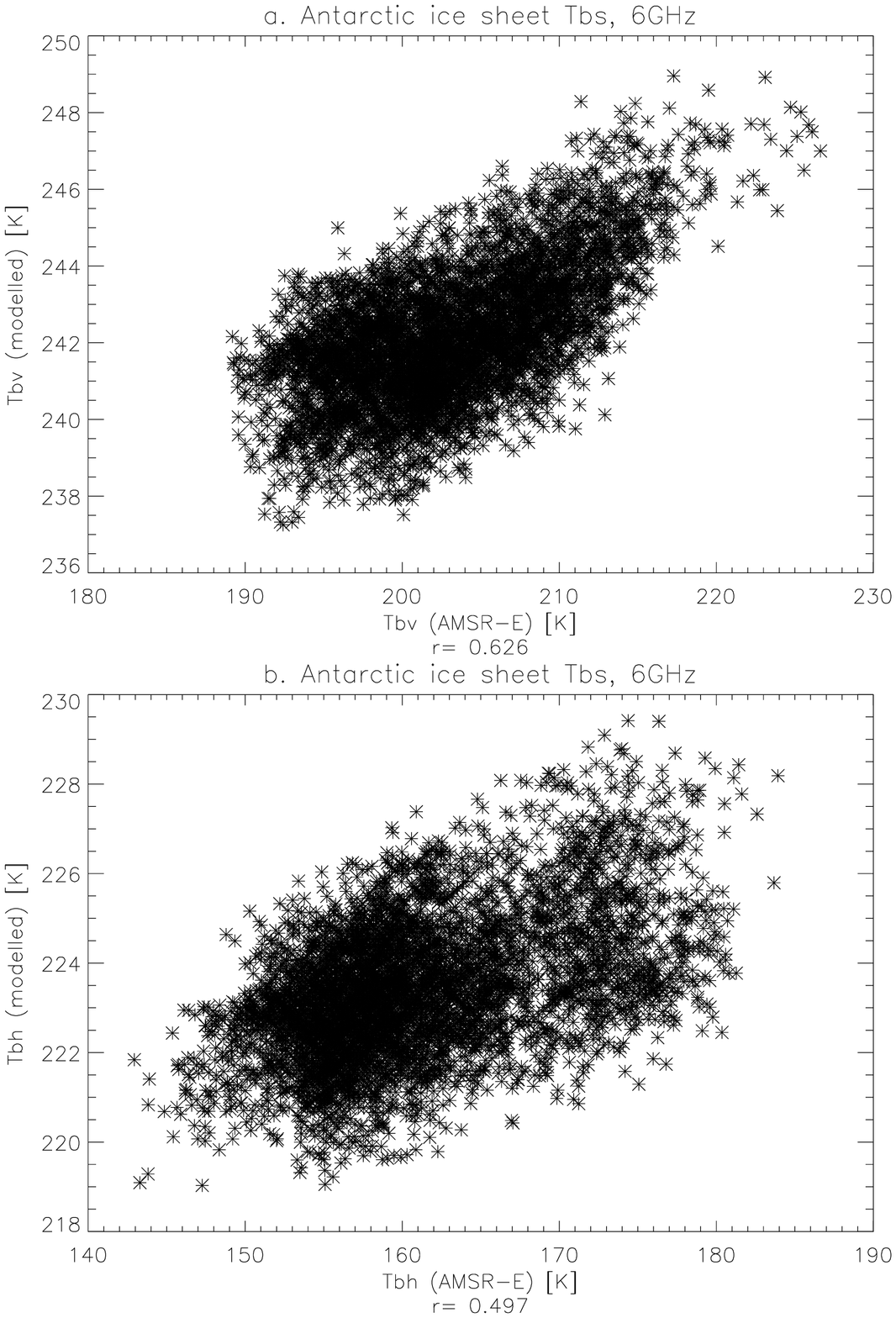}
\caption{Radiative transfer simulations of AMSR-E brightness temperatures
over Antarctic ice pack for the 6 GHz channel.}
\label{icesheet_rt1}
\end{figure}
\begin{figure}
\includegraphics[width=0.9\textwidth]{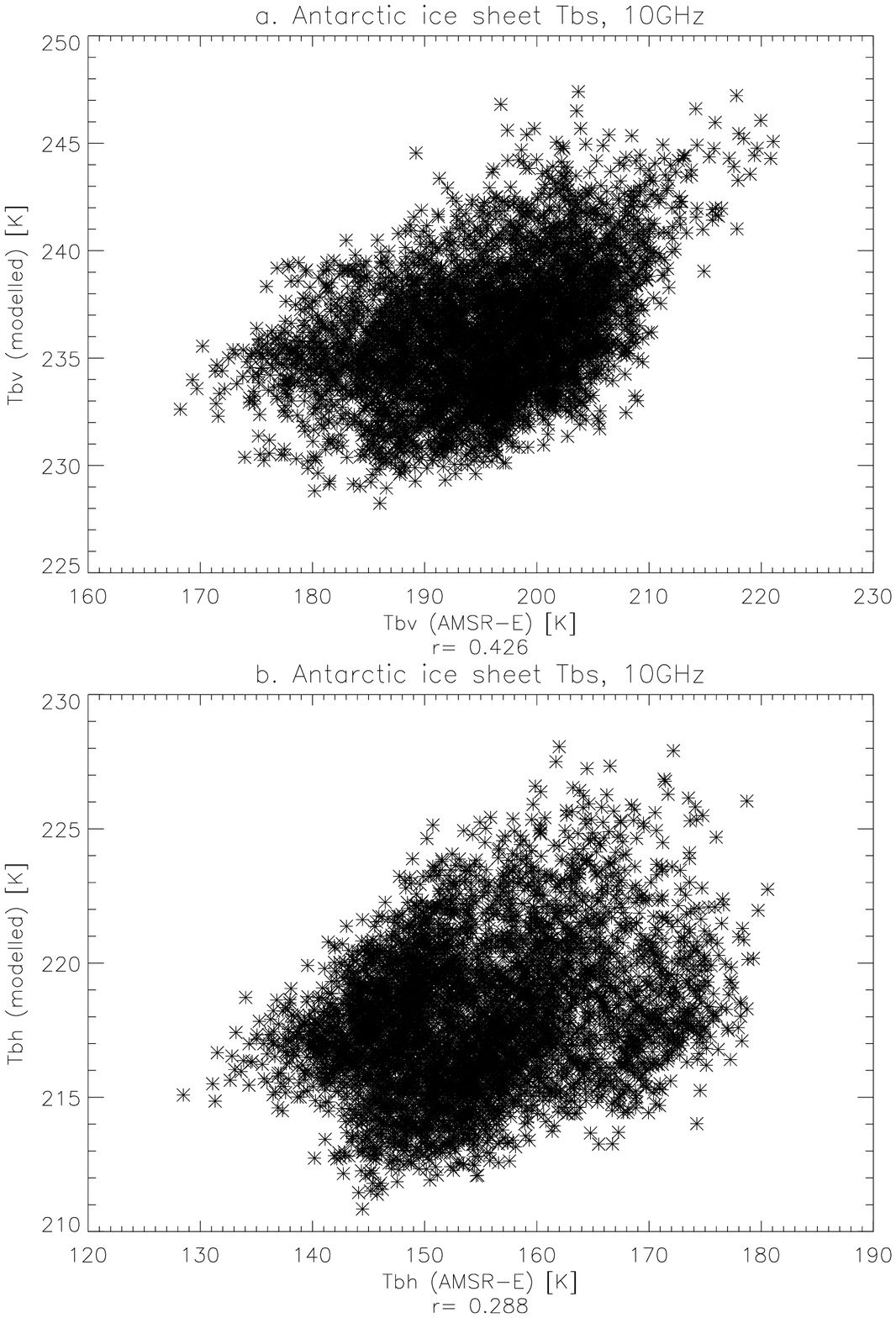}
\caption{Radiative transfer simulations of AMSR-E brightness temperatures
over Antarctic ice pack for the 10 GHz channel.}
\label{icesheet_rt2}
\end{figure}
\begin{figure}
\includegraphics[width=0.9\textwidth]{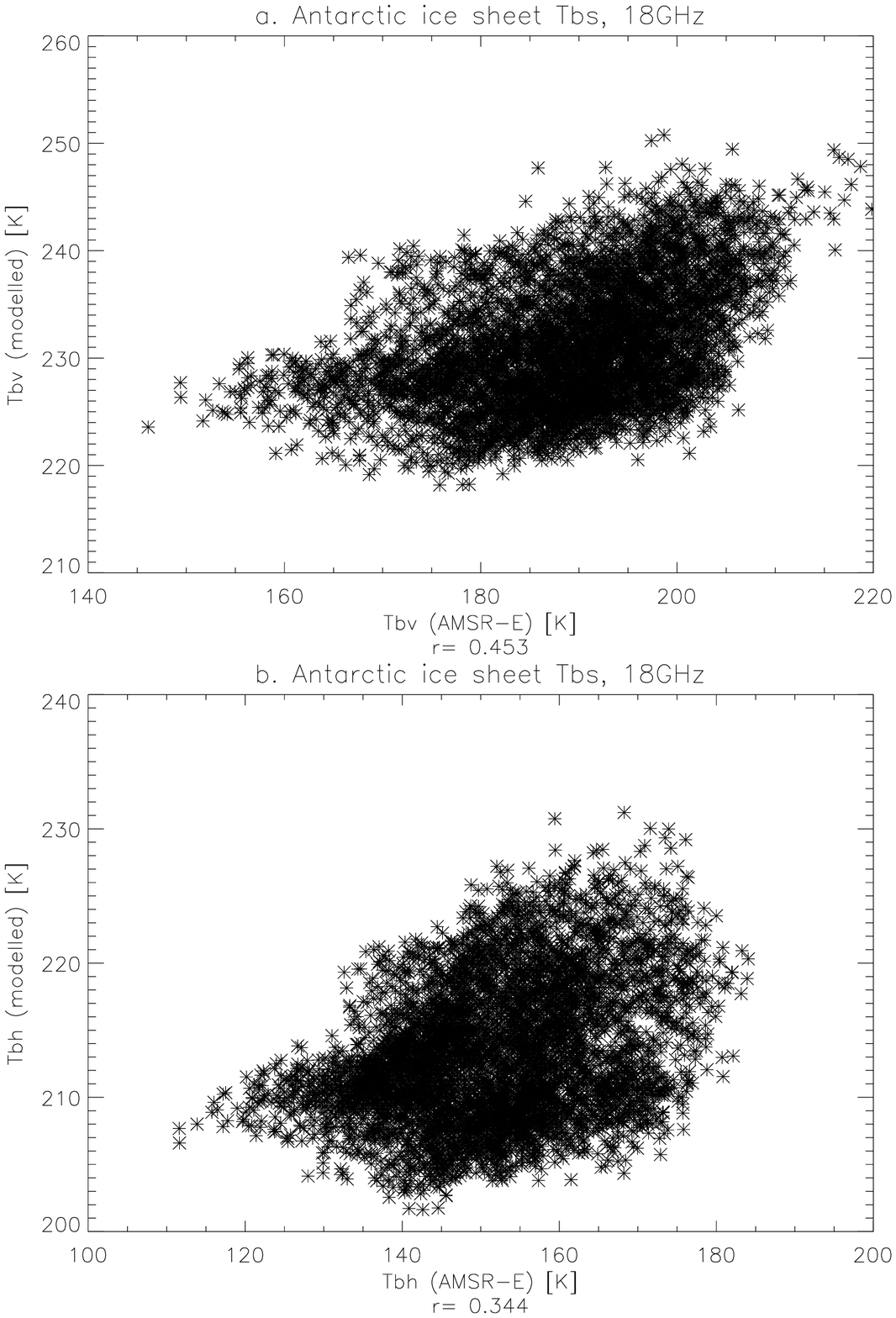}
\caption{Radiative transfer simulations of AMSR-E brightness temperatures
over Antarctic ice pack for the 18 GHz channel.}
\label{icesheet_rt3}
\end{figure}

Icepack simulations are a bit more involved because the density of
the ice varies with depth and because we have to model the internal
temperature using the thermodynamice model described in 
Section \ref{thermo_model}.  5000 measurements were chosen at random
from AMSR-E swath data in a broad area between 0 and 120 degrees East 
longitude 75 and 85 degrees South latitude for the year of 2007.
A thermodynamic model forced by four years of NCEP data was run for
each of the points and then fed to the radiative transfer model.
A sample temperature profile is shown in Figure \ref{sample_tprof}.
Results are shown in Figures \ref{icesheet_rt1} through \ref{icesheet_rt3}.

\section{Discussion and conclusions}

The plane-parallel radiative transfer model is able to simulate radiance
measurements from the AMSR-E satellite instrument well, but with 
large bias.
For the case of lake ice, brightness temperatures are under-estimated for the
6 GHz channels, slightly over-estimated for the 10 GHz vertical channel 
slightly under-estimated for the 10 GHz horizontal channel and badly
over-estimated for the 18 GHz channels, especially in the vertical
polarization.
For higher-frequencies channels, discrepencies are easily accounted for by
the lack of scattering in the model.
As you move up the microwave spectrum from low to high frequency, 
simple absorption and reflection processes become less and less important, 
while scattering begins to dominate.
The tendency of scattering, especially within the ice sheet (as opposed to
surface scattering) is to lower the emissivity along with a complementary
increase in the polarization difference.
Indeed, the measured signal shows an increase
in the polarization difference which is absent from the modelled signal.
In fact, the polarization difference over the ice 
is larger for the measured signal over all frequencies.
This may be accounted for in part by the presence of clouds which
were not screened for when selecting measurement pixels.

Similar discrepencies show up in the open-water signal.
In addition to scattering caused by clouds and rough surface conditions,
the difference may also be caused by water temperatures above freezing
as the temperature of the lake-water was not varied.
Sadly, these are still not point-by-point validations and with only
six data points provide only weak validation of the plane-parallel
RT model.
Recently, \citet{Kang_etal2010} have derived
ice thickness from AMSR-E measurements over the
Great Bear and Great Slave Lakes.
This is a statistical model, however, so it would be good to confirm
the relationship based on more physical considerations,
hopefully paving the way for more physically-based ice retrieval
methods over the ocean, just as exist for retrieval of atmospheric
parameters.  Preliminary results suggest that, except for melting ice,
 the RT model does an
excellent job of predicting AMSR-E radiances based strictly on thickness.
As in the current work, ice temperature was held constant.
Since the lake ice thicknesses in \citet{Kang_etal2010}
were generated using a thermodynamic ice growth model,
temperatures within the ice sheet were returned as a by-product.
The author hopes to publish a more complete study using these to
drive the emissivity model.

\begin{table}
\caption{Biases of the radiative transfer emissivity model compared with
AMSR-E data over Antarctic icepack.  
Linear-regression slope and offset parameters are presented
for all six channels.}\label{rt_stats}
\begin{center}
\begin{tabular}{|c|cc|}
\hline
Channel & Slope & Bias \\
\hline \hline
6v & 0.180 & -39.2 \\
6h & 0.106 & -62.2 \\
10v & 0.157 & -40.7 \\
10h & 0.086 & -63.3 \\
18v & 0.239 & -42.3 \\
18h & 0.148 & -61.4 \\
\hline
\end{tabular}

\end{center}
\end{table}

For a truly point-by-point analysis, we turn to the model of 
Antarctic icepack.
Here the magnitudes are far off, with the model severely over-estimating
the actual values, however the correlation, while not excellent, is
nonetheless strong and significant.
As before, the bias is easily explained by scattering as glacial icepack
tends to be very granular with grain sizes as large as 15 mm
\citep{Rist_etal2002}.
Thus scattering will strongly affect even the lowest, 6 GHz channel.
Biases are shown in Table \ref{rt_stats} which lists regression slope and
offset values for all six channels.
As expected, biases are different for the vertical versus horizontal channels
in keeping with the scattering explanation.
Considering that the only input parameters varied in the emissivity model were
the temperatures, and these were derived indirectly from a thermal
conductivity model driven by reanalysis data, the results seem quite good.

\begin{figure}
\includegraphics[width=0.9\textwidth]{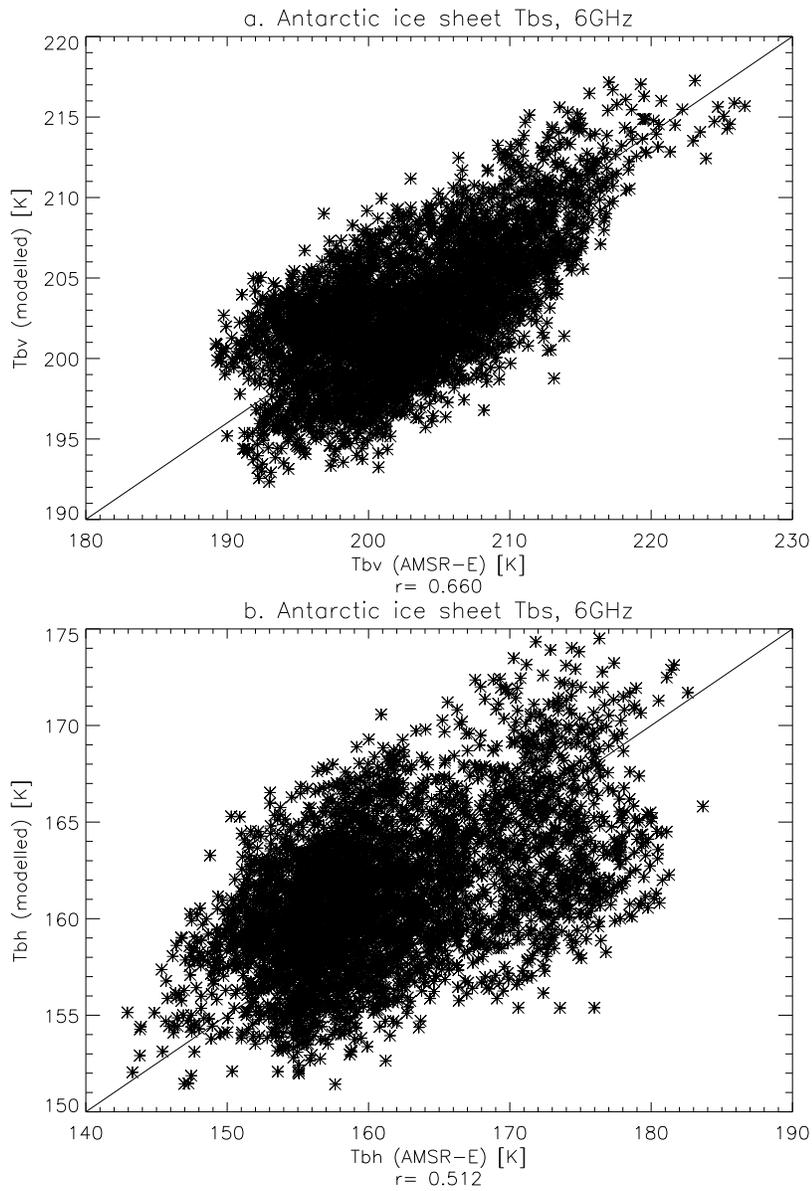}
\caption{Statistical model of AMSR-E brightness temperatures
over Antarctic ice pack for the 6 GHz channel.}
\label{icesheet_stat1}
\end{figure}
\begin{figure}
\includegraphics[width=0.9\textwidth]{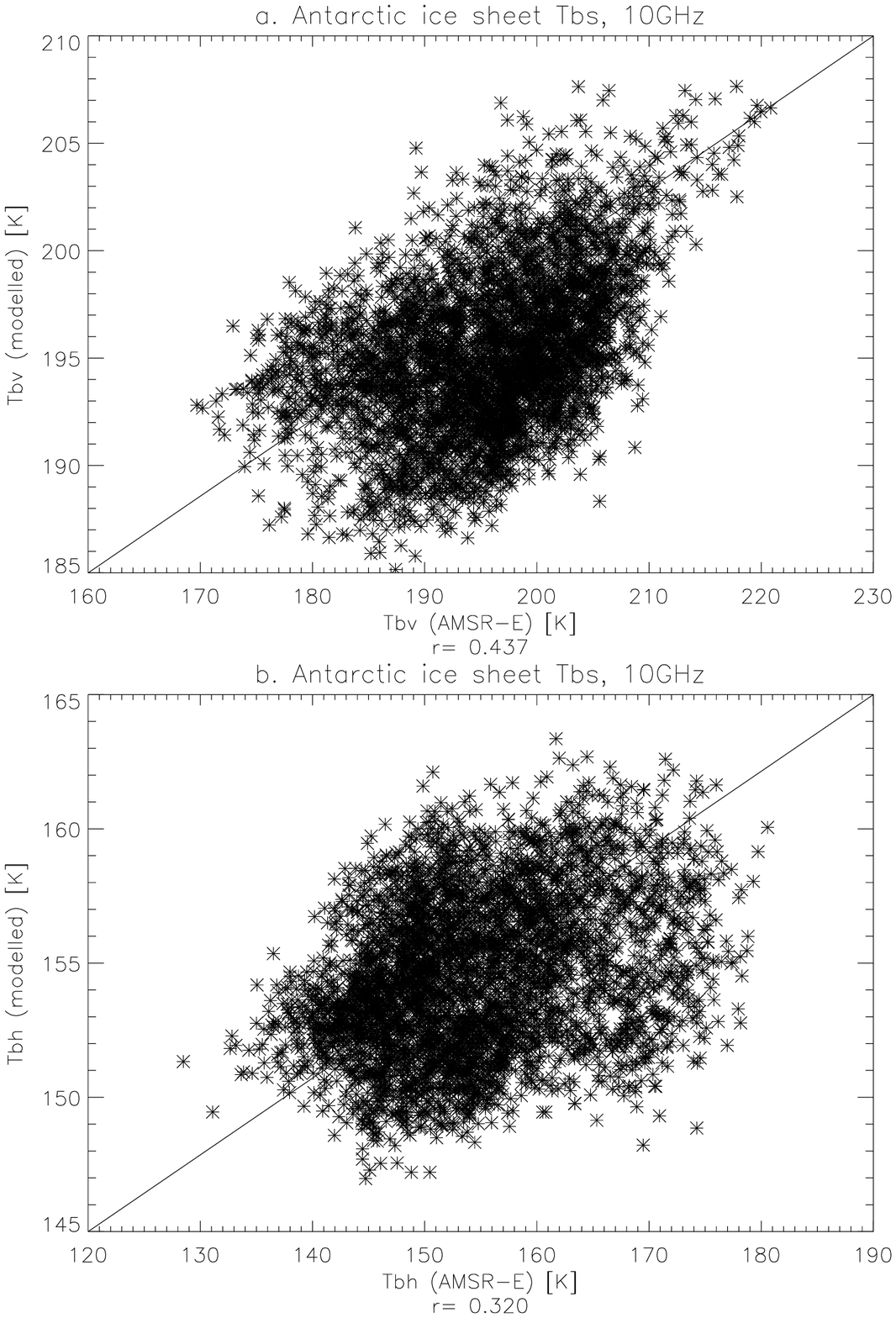}
\caption{Statistical model of AMSR-E brightness temperatures
over Antarctic ice pack for the 10 GHz channel.}
\label{icesheet_stat2}
\end{figure}
\begin{figure}
\includegraphics[width=0.9\textwidth]{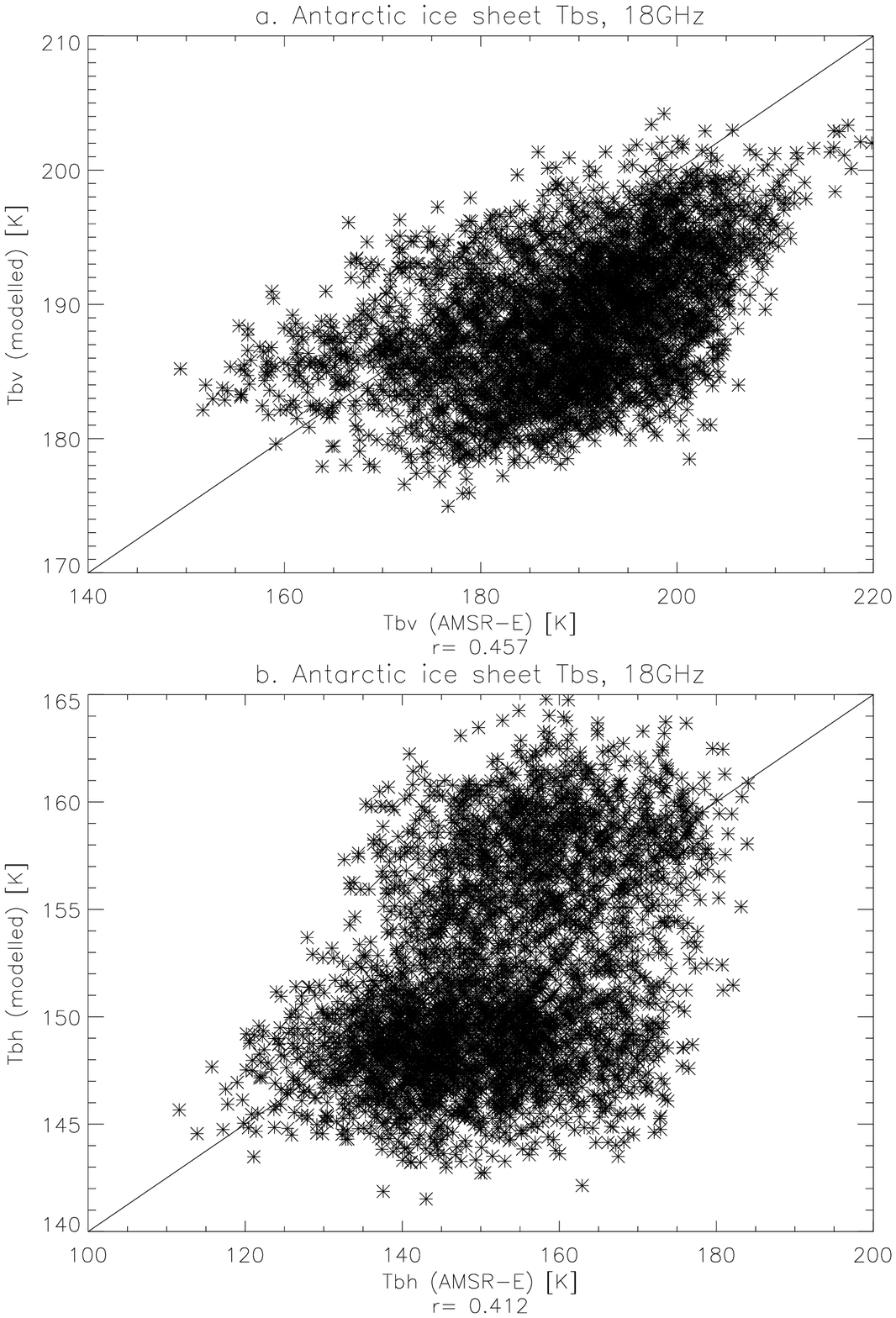}
\caption{Statistical model of AMSR-E brightness temperatures
over Antarctic ice pack for the 18 GHz channel.}
\label{icesheet_stat3}
\end{figure}

\begin{figure}
\includegraphics[width=0.45\textwidth]{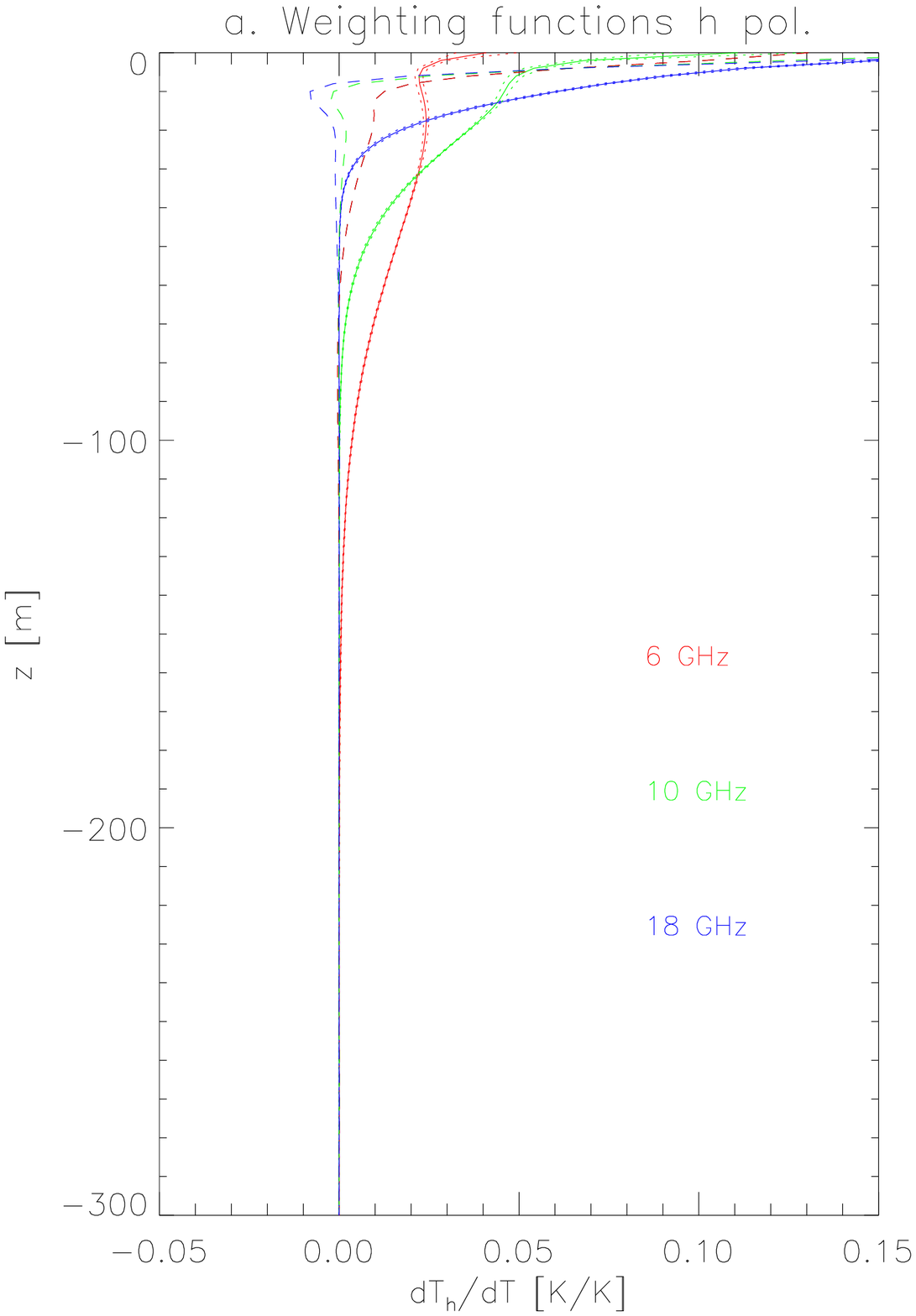}
\includegraphics[width=0.45\textwidth]{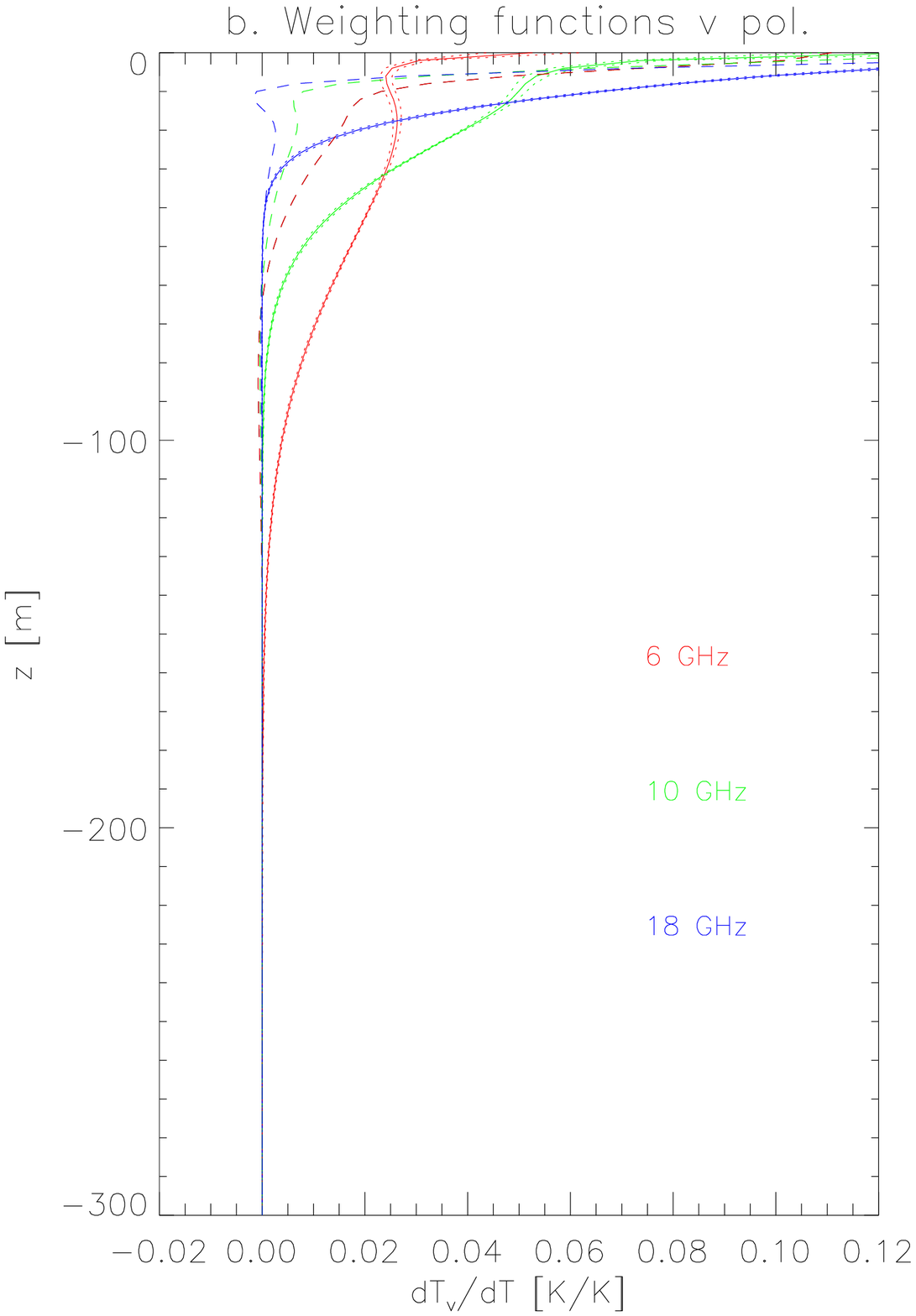}
\caption{Temperature weighting functions for Antarctic ice sheet for horizontal (a) and vertical polarizations (b).  
The weighting functions for the radiative transfer simulations represent 
an average over fifty (50) AMSR-E measurement pixels randomly selected from the test set.
The averages are shown by the solid line while the dotted lines enclose the standard deviations.
The broken lines show the weighting coefficients for the statistical model.}
\label{icewt}
\end{figure}

It is instructive to compare the RT results with a purely statistical model.
Therefore, 1000 temperature profiles were selected for use as ``training data''
and correlated with the AMSR-E brightness temperatures using the top ten (10)
singular vectors.
Predicted brightness temperatures for the other 4000 measuremnt pixels are
presented in Figures \ref{icesheet_stat1} through \ref{icesheet_stat3}
along with AMSR-E brightness temperatures for comparison.
Although the large biases are absent,
the results are remarkable both for the similarities in accuracy as
well as in the shape of the scatterplots.
This is hardly surprising since the RT model is practically linear,
with the only non-linearities arising from the weak dependence of real
permittivity on temperature.

The weighting coefficients for the RT versus statistical model are compared
in Figure \ref{icewt}(a) and \ref{icewt}(b).  For the statistical model,
simulated brightness temperatures are simply a linear combination of the
temperatures of the icepack.  To account for the weak nonlinearity in the
RT model, weights are calculated with a numerical derivative, 
$\frac{\partial T_{\mathrm b}}{\partial T_i}$, averaged over 500 
trials: standard deviations are also shown in the figures.

The two types of weighting functions are quite similar, although the 
statistical weighting functions  are shifted considerably upwards.  
They also allow for negative weight.
Thus, scattering means that the ice pack is effectively much more opaque than our
non-scattering model suggests.
This provides a further explanation for the too-high brightness temperatures
returned by the RT model since ice temperatures tend to be higher further
down in the icepack--see Figure \ref{sample_tprof}.
Nonetheless, the penetration depth of the microwaves is quite deep:
as much as 50 m for the lowest frequency.
This means that radiometer measurements of glacial icepack return information
about temperatures within the ice sheet, which are reflective of historical
temperature.

The interaction of electro-magnetic radiation with matter is a very fundamental
problem.
Most operational ice retrieval algorithms are based primarily on empirical 
parameterizions.
While the model under discussion is somewhat ad-hoc, nonetheless it is based
on real physics.
Such physical models are invaluable for ice retrieval efforts,
both for use as forward models and to retrieve physical characteristics
of ice sheets.
Most likely parameters to be retrieved include ice temperature, ice thickness
and in particular, complex permittivities within the ice sheet which
are reflective of the ice composition and micro-structural properties.
Retrieving fresh-water ice thickness appears to be quite easy,
however retrieving salt-water ice thicknesss is proving to be quite 
difficult \citep{Mills_Heygster2009, Mills_Heygster2011}.

Ice, particularly saline ice, is a highly complex composite.
Because of this complexity, determining its electromagnetic 
properties--how it interacts with radiation--is very difficult.  
More work needs to be done to understand how ice emits, absorbs and
scatters electromagnetic radiation.

\bibliography{fresh_water}

\end{document}